\documentclass[fleqn,10pt]{wlscirep}
\usepackage[utf8]{inputenc}
\usepackage[T1]{fontenc}
\title{Using all-sky differential photometry to investigate how nocturnal clouds darken the night sky in rural areas}

\author[1,2,*]{Andreas Jechow}
\author[1]{Franz H{\"o}lker}
\author[2,1]{Christopher C. M. Kyba}
\affil[1]{Leibniz Institute for Freshwater Ecology and Inland Fisheries, Ecohydrology, M{\"u}ggelseedamm 310, 12587 Berlin, Germany}
\affil[2]{GFZ German Research Centre for Geosciences, Remote Sensing, Telegrafenberg, 14473 Potsdam, Germany}

\affil[*]{andreas.jechow@gmx.de}

\begin{abstract}
Artificial light at night has affected most of the natural nocturnal landscapes worldwide and the subsequent light pollution has diverse effects on flora, fauna and human well-being. To evaluate the environmental impacts of light pollution, it is crucial to understand both the natural and artificial components of light at night under all weather conditions. The night sky brightness for clear skies is relatively well understood and a reference point for a lower limit is defined. However, no such reference point exists for cloudy skies. While some studies have examined the brightening of the night sky by clouds in urban areas, the published data on the (natural) darkening by clouds is very sparse. Knowledge of reference points for the illumination of natural nocturnal environments however, is essential for experimental design and ecological modeling to assess the impacts of light pollution. Here we use differential all-sky photometry with a commercial digital camera to investigate how clouds darken sky brightness at two rural sites. The spatially resolved data enables us to identify and study the nearly unpolluted parts of the sky and to set an upper limit on ground illumination for overcast nights at sites without light pollution.
\end{abstract}
\begin{document}

\flushbottom
\maketitle

\thispagestyle{empty}

\section{Introduction}
Daily, lunar and seasonal cycles of natural light have been key forms of environmental information for organisms since shortly after the first emergence of life \cite{kronfeld2003partitioning}. Artificial light at night (ALAN) has rapidly increased since the 19th century and has been introduced outdoors in places, at times, spectra and intensities at which it does not naturally occur at night \cite{Gaston:2015_ptb,Bara2016}. Global simulations of skyglow (one form of light pollution) on clear nights demonstrate the dramatic extent to which artificial night sky brightness (NSB) has altered nightscapes worldwide \cite{falchi2016WA}, with for example 88\% of the EU experiencing light pollution on clear nights. We have recently shown that ALAN is growing in area and intensity by more than 2$\%$ per year on a global scale, with individual countries having increases of more than 10$\%$ per year \cite{kyba2017VIIRS}.

ALAN thus has become an anthropogenic stressor - light pollution, that can affect humans \cite{cho2015effects}, animals \cite{book:rich_longcore, Longcore2004}, plants \cite{bennie2016plants} and microorganisms \cite{Hoelker:2015}. Recent findings include disruption of pollination \cite{knop2017ALAN} or the disturbance of reproduction of mammals \cite{robert2015artificial}. Despite the many studies regarding the impact of ALAN on the environment, a comprehensive quantitative understanding of the amount of artificial and natural light in the nocturnal environment for all weather conditions and how it is perceived by organisms is still lacking. In particular, while reference values for sky luminance and illuminance exists for clear nocturnal skies \cite{falchi2016WA}, these reference points have not been defined for cloudy conditions. However, knowledge of such reference points is needed for the design of experiments and for ecological modeling to fully evaluate the environmental impacts of ALAN \cite{book:rich_longcore, Longcore2004}.

Atmospheric conditions can change light levels rapidly during the day \cite{wacker2015cloud} and the night \cite{Kyba:2011_sqm}. In a natural setting with no ALAN, clouds should darken the night sky in most cases. For large distances around today's urban areas however, clouds usually amplify the effects of ALAN, creating a complete reversal of natural conditions \cite{Kyba:2011_sqm, Kyba:2012_mssqm, Kyba:2015_isqm, Ribas2016clouds, jechow2017balaguer}. NSB data on rural sites with low ALAN levels, however, is sparse and in general few studies with imaging devices and clouds exist.

One reason for these knowledge gaps in light pollution research is due to the fragmentation of expertise across very different disciplines and the lack of widely applicable commercial measurement tools until recently. For decades, NSB monitoring was mainly of interest for astronomers as they were first to study light pollution \cite{Rosebrugh:1935, Walker:1970, riegel1973, Bertiau:1973}. For logical reasons, astronomers mainly concentrated their research on clear skies, but also performed comprehensive observations at several important (often very remote) astronomical sites, for example at high altitudes \cite{patat2008dancing, plauchu2017night, aube2014evaluation} or latitudes \cite{yang2017optical}. The advent of small photometers customized for NSB measurements (mainly the Sky Quality Meter: SQM) has sparked NSB research beyond professional astronomy. Nowadays scientists from different fields, amateur astronomers, and citizen scientists obtain NSB data from local to global scale \cite{Bara2016, falchi2016WA, Kyba:2013_GaN, Kyba:2015_isqm}. However, these single channel sensors usually provide zenith NSB data in a single spectral band, which provides only limited information of the nocturnal night light environment and might require cross-calibration \cite{fruck2015instrumentation, RibasLonne2017}. Commercial digital cameras with fisheye lenses are a promising tool for all-sky nighttime photometry to fill this gap, yielding spatially resolved NSB data of nearly the full hemisphere in three spectral channels \cite{Kollath:2010,solano2016allsky,kollath2017night, jechow2017measuring,jechow2017balaguer}. A more comprehensive discussion of state-of-the-art NSB measurement techniques can be found in a recent review \cite{haenel2018measuring}.

Recently, we reported on lowered zenith luminance values for cloudy nights using an SQM at Lake Stechlin, Germany \cite{Jechow2016} while others observed both, brightening and darkening by clouds in SQM datasets at Montsec Astronomical Park, Spain \cite{Ribas2016clouds} and in Austria \cite{Posch2018}. All-sky imaging data of cloud amplification were investigated in Madrid, Spain \cite{bara2015zernike} and near Montsec Astronomical Park, Spain \cite{jechow2017balaguer}. However, imaging data on the darkening by clouds in the context of ALAN has not been published, yet. To set a lower limit for nighttime luminance and illuminance such data is essential as it provides information of the NSB across the full hemisphere. 

In this work, we use differential all-sky photometry \cite{jechow2018differential} with a commercial DSLR camera to investigate the changes that clouds have upon the NSB of both the entire sky, and segments of the sky. We examine two rural locations: one 70 km north of Berlin, Germany (LakeLab, Lake Stechlin, Brandenburg) and one 180 km east of Cape Town, South Africa (Night Sky Caravan Farm, Bonnievale). The calculated all-sky luminance maps and the derived color corrected temperature (CCT) from the three spectral channels of the DSLR camera enable us to study the NSB distribution in detail. The spatially resolved data allows to identify and study the minimally polluted parts of the sky for locations that are not completely free of artificial skyglow. Our results fill a research gap in the context of ecological light pollution and we hope this work triggers more such spatially resolved NSB measurements on a global scale to increase understanding of the impacts of ALAN.

\section{Results}
\subsection{Overcast sky at LakeLab, Lake Stechlin, Germany}
The first observations were taken at the LakeLab, situated about 70 km north of the city of Berlin. Data was taken during a clear night on August 11th 2016 at 01:29 local time and the following overcast night on August 12th 2016 at 01:27 local time (+2 hours GMT). The cloud base height for the overcast night was inferred to be $h_{CB} \approx$ 850 m $\pm$ 200 m. More information on the site and the surroundings is summarized in the methods section.

\subsubsection{Full color images, LakeLab}
\begin{figure}[htbp]
\centering
\includegraphics[width=0.8\columnwidth]{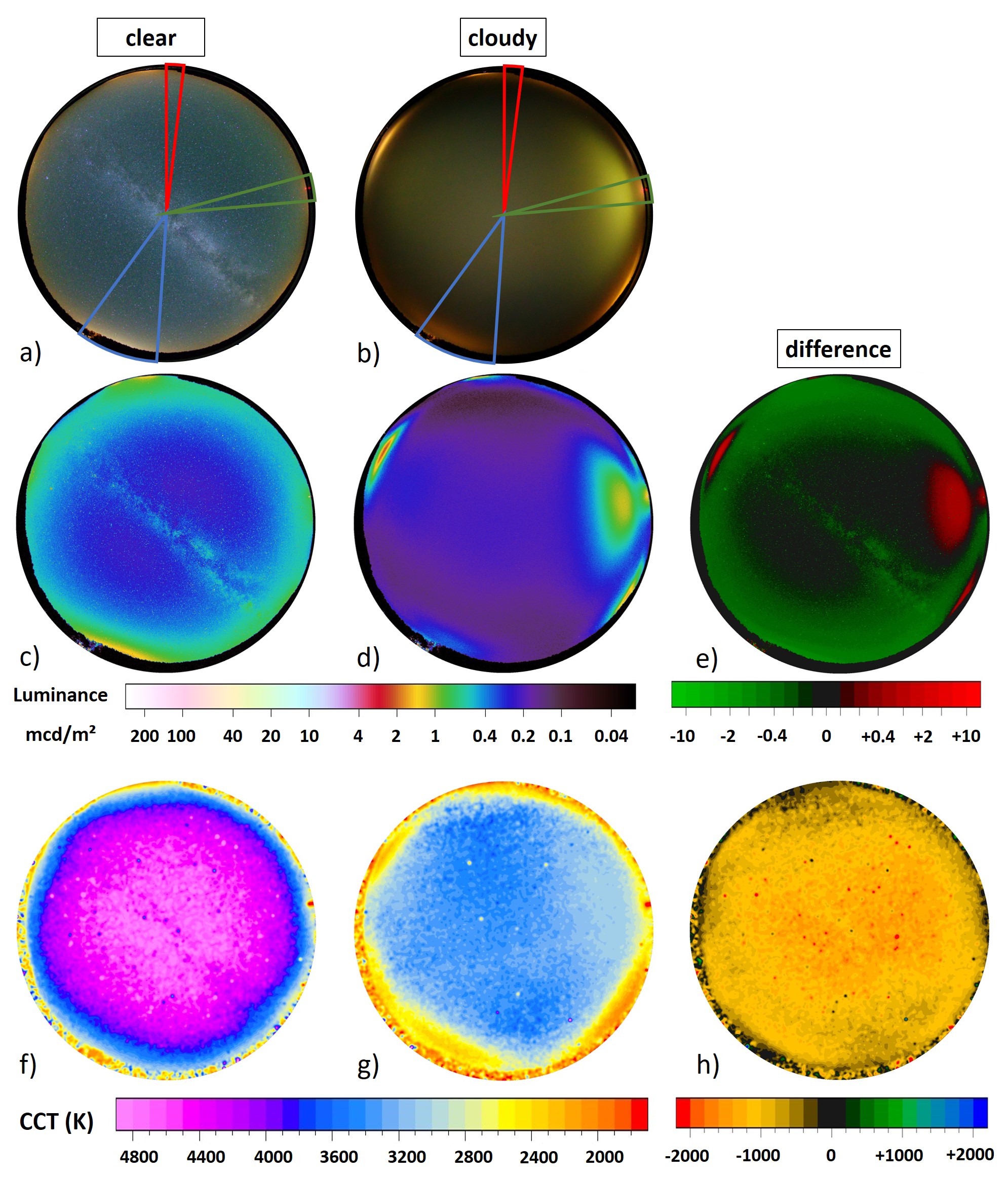}
\caption{Imaging data for LakeLab, Lake Stechlin Germany. The upper row (a,b) shows all-sky RGB images for a) a clear sky on August 11th 2016 at 01:29 local time and b) an overcast sky on August 12th 2016 at 01:27 local time local time. The colored segments are analyzed in more detail in subsection 2.1.2. The middle row (c,d,e) shows all-sky luminance maps calculated from the RGB images for c) a clear night and d) an overcast night. The right hand luminance map (e) shows the difference between the two nights obtained from subtracting images. The lower row (f,g,h) shows all-sky CCT maps calculated from for f) a clear night and g) an overcast night. The right hand CCT map (h) shows the difference between the two nights obtained from subtracting clear night data from the cloudy night data.}
\label{LL_IMG}
\end{figure}

The upper row in Figure \ref{LL_IMG} shows all-sky images in full color (RGB) obtained at LakeLab on Lake Stechlin for a) a clear sky and b) an overcast sky. The upper part of the image is pointing towards north, the lower towards the south, west is on the right-hand side and east on the left-hand side, respectively. In Fig \ref{LL_IMG} a), the Milky Way is prominent and some skyglow is apparent near the horizon. In Fig \ref{LL_IMG} b), the Milky Way is obstructed by clouds, and a bright glow is apparent in the west. A change in color from blue-green (clear sky) to yellow (clouds) is notable. Three regions of interest for further investigation are indicated in Fig. \ref{LL_IMG} a) and b). The red region points to a sky segment with almost no apparent skyglow from nearby settlements, roughly to the north, centered at an azimuth angle of 350$^{\circ}$ spanning over a width of 10$^{\circ}$. The blue region points roughly to the south, centered at an azimuth angle of 160$^{\circ}$ (Berlin direction) spanning over a width of 30$^{\circ}$. The green region points roughly to the west, centered at an azimuth angle of 280$^{\circ}$ (Rheinsberg power plant direction) spanning over a width of 10$^{\circ}$.

\subsubsection{Luminance maps}
From the all-sky images shown in Fig. \ref{LL_IMG} a) and b), the luminance (similar to ''brightness'' with respect to human vision) was calculated using the software Sky Quality Camera (SQC, see methods) for each pixel. The resulting luminance maps are shown in Fig. \ref{LL_IMG} c) for the clear night and d) the overcast night.

For the clear night (Fig. \ref{LL_IMG} c), the luminance ranges from near-natural values of about 0.25 mcd/m$^2$ around the zenith (excluding the Milky Way) to 1.0-2.0 mcd/m$^2$ in the direction of nearby and distant settlements (natural sky is ca. 0.20 - 0.25 mcd/m$^2$ \cite{falchi2016WA}). These include the city of Berlin (azimuth ca. 160$^{\circ}$) in the south-southeast (bottom left) and the towns of Neustrelitz (azimuth ca. 5$^{\circ}$) and Neubrandenburg (azimuth ca. 10$^{\circ}$), both to the north (top of image). There is a small skyglow belt near the horizon with an average luminance reaching values on the order of 0.7 mcd/m$^2$. See Fig. \ref{WA_LL} in the methods section showing maps of the region.

For the overcast night (Fig. \ref{LL_IMG} d), the luminance of 0.20 mcd/m$^2$ near zenith is slightly lower than for the clear night, and the skyglow belt near the horizon has vanished. The horizon is dark in most directions, with the notable exception of bright areas in the direction of nearby light sources. Most striking of these is the former nuclear power plant of Rheinsberg at ca. 2.8 km distance to the west (azimuth ca. 280$^{\circ}$, right hand side). Others are F{\"u}rstenberg (azimuth ca. 60$^{\circ}$) and Rheinsberg (azimuth ca. 240$^{\circ}$). All three local sites are not very apparent in the clear sky data.

The difference in luminance between the two nights was acquired by subtracting the clear night data from the cloudy night data. The result is shown in Fig. \ref{LL_IMG} e), with green regions being darker with clouds and red regions brighter. The subtracted image shows that the decrease in NSB at the zenith is mainly due to the attenuation of the Milky Way's light by clouds. Furthermore, clouds obviously darken the sky near the horizon with the exceptions of the power plant (right side, west) and the two skyglow regions at the very edge of the image (upper right, northeast and lower left, southwest). Strikingly, on clear nights the skyglow from distant sources (e.g. from Berlin, Neustrelitz, Neubrandenburg) is dominant, while the overcast skyglow is dominated by local light sources (Power plant, Rheinsberg and F{\"u}rstenberg).

\subsubsection{CCT maps}
The lower row in Fig. \ref{LL_IMG} (f,g,h) shows CCT maps calculated from the RGB images at LakeLab. Fig. \ref{LL_IMG} f) shows the clear night and g) the overcast night data, while Fig. \ref{LL_IMG} h) shows the difference where the data from the clear night was subtracted from the cloudy night. During the clear night (Fig. \ref{LL_IMG} f), the CCT near zenith (Milky Way up) is 4700 K and the CCT values decrease towards the horizon. For the overcast night (Fig. \ref{LL_IMG} g), the CCT values around zenith (10$^{\circ}$ circle) are on the order of 3400 K. The skyglow regions to the south east, north east and south west reach low CCT values on the order of 2000 K to 2500 K. The skyglow from the power plant (west) has about 3000 K CCT, hardly distinguishable from the background. In the CCT difference map, shown in Fig. \ref{LL_IMG} h), red and yellow colors indicate a decrease in CCT due to clouds, while green and blue regions indicate an increase in CCT by clouds. Overall, only a decrease in CCT by clouds is observed.

\subsubsection{Angular luminance distribution}
\begin{figure}[htbp]
\centering
\includegraphics[width=0.8\columnwidth]{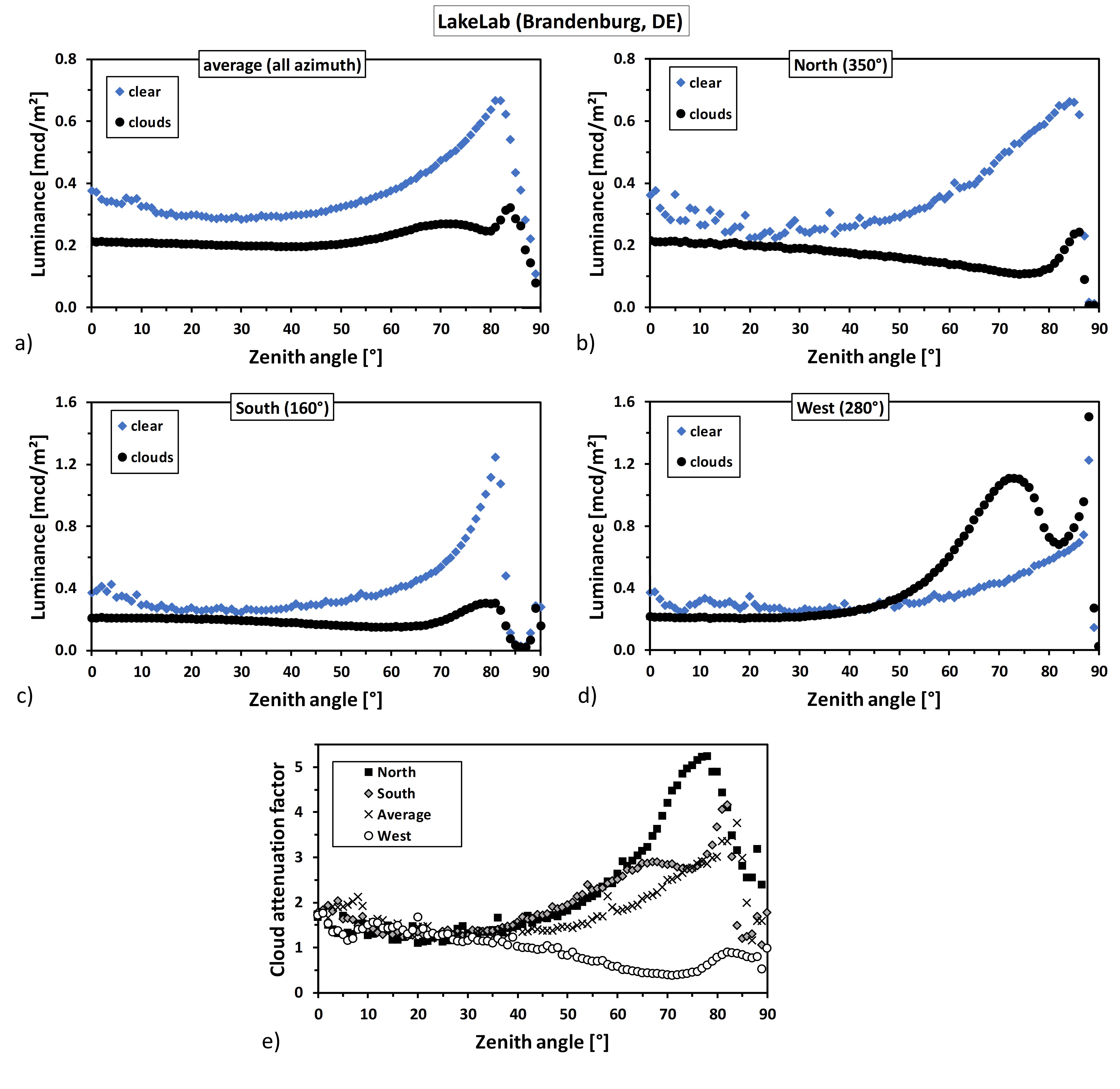}
\caption{The two upper rows show the angular luminance distribution at LakeLab, Germany, for a) the average over all azimuth angles, b) the sky segment facing the north (blue area in Fig. \ref{LL_IMG}) c) the sky segment facing the south (blue area in Fig. \ref{LL_IMG}) and d) the sky segment facing the west (red area in Fig. \ref{LL_IMG}). Each graph shows clear night (blue diamonds) and overcast night (black circles) data. Note the difference in scale between the two upper rows. The variability near zenith in the clear sky data originates from bright stars. The lower row shows cloud attenuation factors at LakeLab, Germany, for each investigated sky segment derived from the angular luminance distribution shown in a)-d) obtained by dividing the clear sky data by the cloudy sky data. Black squares represent the northern segment, diamonds the southern segment, crosses the average and circles the western segment.}
\label{LL_graphs}
\end{figure}

From the spatially resolved data shown in Fig. \ref{LL_IMG} c) and d), we can derive the angular distribution of luminance from zenith to the horizon. This can be done for the whole image, and for segments of certain azimuth angles (see Fig. \ref{LL_IMG} a) and b) and text above. The angular luminance distribution is shown in Fig. \ref{LL_graphs} for a) the average over all azimuth angles, b) the sky segment facing the north c) the sky segment facing the south and c) the sky segment facing the west. Each graph shows clear (blue diamonds) and overcast (black circles) data. The region in the north shown in Fig. \ref{LL_graphs} b) was chosen as it is least affected by skyglow, the one in the south (Berlin direction), shown in Fig. \ref{LL_graphs} c), was chosen as it is most affected by distant skyglow and the one in the west (power plant direction), shown in Fig. \ref{LL_graphs} d), was chosen as it is most affected by nearby skyglow.

All plots show a darkening by clouds near zenith. For the region in the north and the south as well as the average over the whole image, the luminance of the cloudy night is lower than that of the clear night for all zenith angles. For the region in the west, the luminance of the cloudy night is lower only for zenith angles below 40$^{\circ}$.

The general trend for the clear sky data is that the near-zenith values are elevated due to the presence of the Milky Way and that apart from that, the luminance rises towards the horizon. This rise depends on the amount of artificial skyglow present at the individual direction. No such general trend is obvious in the cloudy sky data. However, for the north and south, the luminance of the cloudy sky decreases towards the horizon and then peaks at different zenith angles. Minima occur at about 71$^{\circ}$ for the north reaching 0.11 mcd/m$^2$ and at 60$^{\circ}$ for the south reaching 0.15 mcd/m$^2$. The peaks occur at about 86$^{\circ}$ for the north reaching 0.24 mcd/m$^2$ and at 81$^{\circ}$ for the south reaching 0.30 mcd/m$^2$, respectively. 

The decrease in luminance towards the horizon on the cloudy night is almost overcompensated for the direction towards the west. However, luminance values remain almost constant between 0.20 and 0.21 mcd/m$^2$ up to 30$^{\circ}$. A distinct peak with a maximum luminance of 1.1 mcd/m$^2$ is apparent at a zenith angle of 72$^{\circ}$. We attribute this peak to the lighting system of the former nuclear power plant of Rheinsberg, which is not visible in the clear sky data.

\subsubsection{Cloud attenuation factors}
We calculated cloud attenuation factors for each investigated sky segment from the angular luminance distribution shown in Fig. \ref{LL_graphs} a)-d) by dividing the clear sky data by the cloudy sky data. The result is plotted in the lower row of Fig. \ref{LL_graphs} e). Black squares represent the northern segment, diamonds the southern segment, crosses the average and circles the western segment. Near zenith, a darkening by almost a factor of 2 is observed. Above 30$^{\circ}$, the western segment shows a brightening by clouds, while the other show darkening. In the north (black squares) a maximum attenuation factor of 5.2 is observed at 78$^{\circ}$, in the south a maximum attenuation factor of 4.2 is observed at 82$^{\circ}$ and in the averaged data, a maximum attenuation factor of 3.8 is observed at 84$^{\circ}$. In the west the attenuation factor reaches a minimum of 0.4 (i.e. a brightening factor of 2.5) at 72$^{\circ}$.

\subsubsection{Illuminance}
Horizontal and scalar illuminance (the total luminous flux incident on a surface, see methods) is calculated by the SQC software by integrating over the whole image. It is further possible to estimate (a hypothetical) illuminance assuming a certain angular luminance distribution for zenith or azimuth angles. For example, it is common to assume a constant luminance for all azimuth and zenith angles to estimate the illuminance from zenith measurements taken with single channel devices like an SQM (see methods). We can further assume a non homogeneous luminance distribution varying with zenith angle, but with a uniform value for all azimuth angles. In that case, the average luminance distribution shown in Fig. \ref{LL_graphs} a) would produce equal illuminance as that obtained via integration over all data points. Furthermore, we can infer a hypothetical illuminance under minimal skyglow conditions by assuming that the angular luminance distribution of the northern sky segment (least affected by skyglow) would be valid for all azimuth angles. In table \ref{table_ill_LL}, the horizontal and scalar illuminance from the full image is compared to the inferred illuminance values from the northern sky segment and for illuminance inferred from an SQM handheld photometer (see methods).
\begin{table}[h]
  \centering
  \caption{Illuminance values calculated from the angular luminance distribution for a clear and overcast night at LakeLab, Germany. (*) hypothetical value assuming an equal angular luminance distribution for all azimuth angles (**) inferred value assuming a constant luminance for all azimuth and zenith angles. Zenith luminance from SQM and DSLR are given as reference.}
  \label{table_ill_LL}
  \begin{tabular}{ccc}
	\hline
&clear & clouds\\

horizontal Illuminance [mlx] \\
    \hline
average (full data) & 1.10 $\pm$ 0.11 & 0.68 $\pm$ 0.07\\
northern segment (*)  & 1.04 $\pm$ 0.10 & 0.51 $\pm$ 0.05\\
SQM (**) & 0.92 $\pm$ 0.09 & 0.69 $\pm$ 0.07\\
\hline
\\
scalar Illuminance [mlx] \\
    \hline
average (full data) & 2.44 $\pm$ 0.24 & 1.42 $\pm$ 0.14\\
northern segment (*)  & 2.33 $\pm$ 0.23 & 0.92 $\pm$ 0.09\\
SQM (**) & 1.84 $\pm$ 0.18 & 1.38 $\pm$ 0.14\\
\hline
\\
Zenith luminance [mcd/m$^2$]\\
\hline
SQM & 0.29 $\pm$ 0.03 & 0.22 $\pm$ 0.02\\
DSLR & 0.30 $\pm$ 0.03  & 0.21 $\pm$ 0.02 \\
\hline
  \end{tabular}
\end{table}

\subsection{Cloudy sky in remote area near Cape Town, South Africa (Night Sky Caravan Park, Bonnievale)}
The second set of observations were obtained at Night Sky Caravan Farm near Bonnievale, South Africa for a nearly cloud-free sky on March 15th 2018 at 22:44 local time and a cloudy sky on March 14th 2018 at 21:50 local time (+2 hours GMT). The cloud base height for the cloudy night was inferred to be $h_{CB} \approx$ 1.5 km $\pm$ 0.2 km. The site lies in the Cape Wine district about 180 km east of Cape Town (see methods for details).

\begin{figure}[htbp]
\centering
\includegraphics[width=0.8\columnwidth]{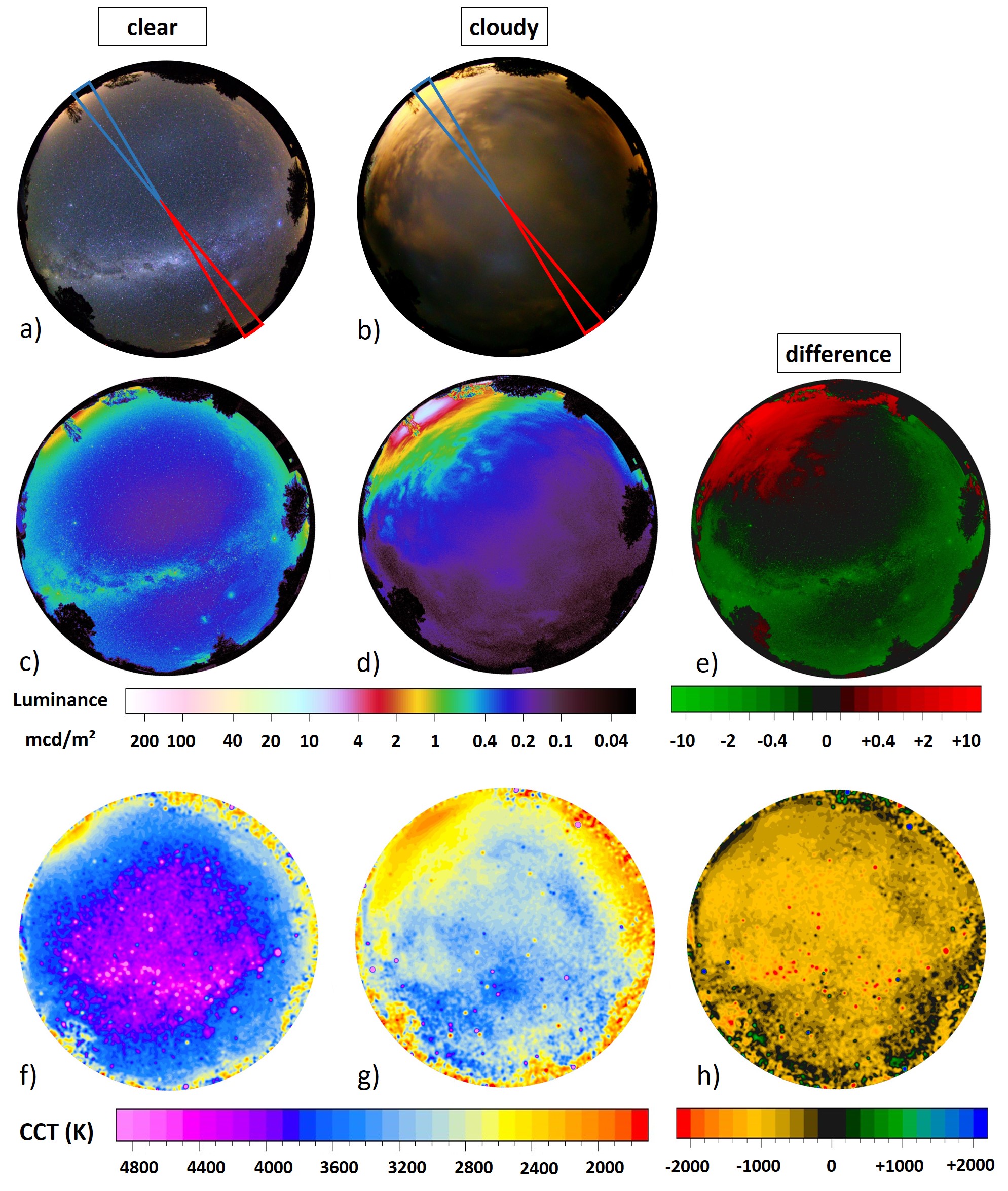}
\caption{The upper row shows all-sky RGB images obtained at Night Sky Caravan Park, near Bonnievale, South Africa for a) a clear sky on March 15th 2018 at 22:44 local time and b) a cloudy sky on March 14th 2018 at 21:50 local time. The colored segments are analyzed in more detail in subsection 2.2.2. The middle row all-sky luminance maps calculated from the RGB images for c) the clear night and d) an overcast night. The right hand luminance map (e) shows the difference between the two nights obtained from subtracting images. the lower row shows all-sky maps of CCT calculated from the RGB images for f) a clear night and g) an overcast night. The right hand CCT map (h) shows the difference between the two nights obtained from subtracting clear night data from the cloudy night data.}
\label{ZA_IMG}
\end{figure}

\subsubsection{Full color images, Bonnievale}
The upper row in figure \ref{ZA_IMG} shows all-sky RGB images obtained at Night Sky Caravan Park near Bonnievale, South Africa for a) an almost clear sky and b) a cloudy sky. The upper part of the image is pointing towards north, the lower towards the south, west is on the right-hand side and east on the left-hand side, respectively. In Fig \ref{ZA_IMG} a), the well structured Milky Way is prominent and some skyglow is apparent near the horizon, particularly in the northeast. Some small patchy clouds can be seen in the eastern direction. In Fig \ref{ZA_IMG} b), the Milky Way is obstructed by clouds and a bright glow is apparent in the northwest. The southeastern direction however, is very dark. As for the photos from Germany, a change in color from blueish (clear sky) to yellowish (clouds) is also notable.

Two regions of interest for further investigation are indicated in Fig. \ref{ZA_IMG} a) and b). The red region points to a region with almost no apparent skyglow from nearby settlements, roughly to the south-west, centered at an azimuth angle of 216$^{\circ}$ spanning over a width of 10$^{\circ}$. The blue region points roughly to the region with the highest skyglow near the horizon, centered at an azimuth angle of 36$^{\circ}$ (Bonnievale direction) spanning over a width of 10$^{\circ}$.

\subsubsection{Luminance maps}
The middle row in Fig. \ref{ZA_IMG} (c,d,e) shows luminance maps c) for the clear night and d) the overcast night, whereas e) shows the difference from subtracted data. For the clear night (Fig. \ref{ZA_IMG} c), the luminance ranges from near-natural values at zenith of about 0.20 mcd/m$^2$ around the zenith (Milky Way is not at zenith) to 2.0 mcd/m$^2$ in the north-eastern direction. This skyglow originates from the small town of Bonnievale (12 km distance). Skyglow from other nearby towns like Robertson (25 km north), Swellendam (40 km east) or Worcester (65 km north-west) are not apparent in the luminance map as the skyglow is confined due to the nearby mountain ranges or obstructed by trees. See Fig. \ref{WA_ZA} in the methods section for a map of the region.

For the overcast night (Fig. \ref{ZA_IMG} d), the luminance around the zenith (10$^{\circ}$ circle) is slightly lower than for the clear night, reaching values of about 0.19 mcd/m$^2$. The skyglow of Bonnievale is clearly amplified by clouds, reaching luminance values up to 10 mcd/m$^2$. The sky in the opposite direction, however, is darker with the presence of clouds, reaching minima of about 0.05 mcd/m$^2$.

The difference in luminance between the two nights is shown in Fig. \ref{ZA_IMG} e) with green regions being darker with clouds and red regions brighter with clouds. The skyglow in the north-east is amplified (upper left) and the sky light in the south-west attenuated by the clouds, while the luminance directly at zenith is similar under both conditions.

\subsubsection{CCT maps}
The CCT maps obtained from the RGB data for Bonnievale are shown in Fig. \ref{ZA_IMG} f) for the clear night and g) the overcast night, while Fig. \ref{ZA_IMG} h) shows the CCT difference from subtracted data. During the clear night, the CCT values near zenith (10$^{\circ}$ circle) are on the order of 4000 K, with the Milky Way having slightly higher values on the order of 4500 K compared to the sky background. The CCT values decrease towards the horizon. The skyglow of Bonnievale near the horizon has values of 2200 K. For the overcast night, the CCT values around zenith (10$^{\circ}$ circle) are on the order of 3400 K. Over most of the sky, the CCT values remain relatively constant. The skylow regions to the southeast, northeast and southwest reach low CCT values on the order of 2000 K to 2500 K. The difference in CCT between the two nights is shown in Fig. \ref{ZA_IMG} h), where the data from the clear night was subtracted from the cloudy night. Red and yellow colors indicate a decrease in CCT due to clouds, while green and blue regions indicate an increase in CCT by clouds. Overall, only a decrease in CCT by clouds is observed.

\subsubsection{Angular luminance distribution}
\begin{figure}[htbp]
\centering
\includegraphics[width=0.4\columnwidth]{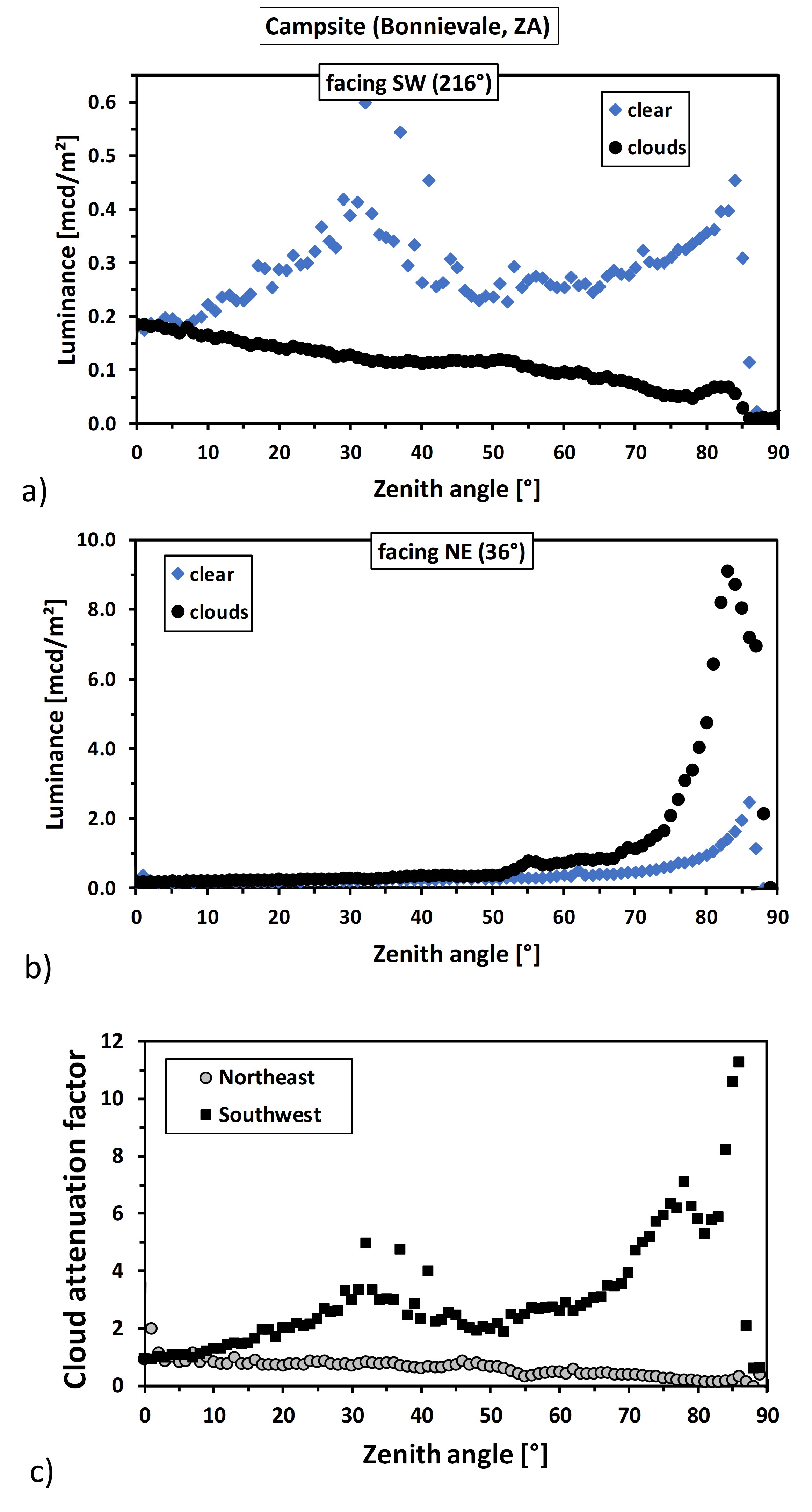}
\caption{The two upper graphs a) and b) show the angular luminance distribution measured near Bonnievale, South Africa, for a), the sky segment facing the south-west (red area in Fig. \ref{ZA_IMG}) and for b) the sky segment facing the north-east (blue area in Fig. \ref{ZA_IMG}). Each graph shows clear night (blue diamonds) and overcast night (black circles) data. The variability slightly off zenith in the clear sky data originates from bright stars. Note the large difference in scale between the two upper graphs. The lower graph shows the cloud attenuation factor, for the two selected sky segments. Grey circles represent the north-eastern segment and black squares the sky segment facing the south-west.}
\label{ZA_graphs}
\end{figure}

The angular luminance distribution is shown in the two upper graphs in Fig. \ref{ZA_graphs} for a), the sky segment facing the south-west (red area in Fig. \ref{ZA_IMG}) b) the sky segment facing the north-east (blue area in Fig. \ref{ZA_IMG}). Each graph shows clear night (blue diamonds) and overcast night (black circles) data.

For the region in the south-west (low skyglow) shown in Fig. \ref{ZA_graphs} a), the luminance of the cloudy night is lower than that of the clear night for all zenith angles. At zenith, the clear sky has near-natural luminance values on the order of 0.2 mcd/m$^2$. The luminance rises with zenith angle, peaking at around 30$^{\circ}$ due to the Milky Way and then rising to about 0.5 mcd/m$^2$ near the horizon. The luminance of the cloudy sky is also close to 0.2 mcd/m$^2$ near the zenith and gradually decreases towards the horizon. The luminance minimum of 0.05 mcd/m$^2$ is reached at about 75$^{\circ}$. For zenith angles above 85$^{\circ}$ the luminance drops to values near zero, as this part of the sky is shielded by nearby objects.

For the region in the north-east (Bonnievale direction) shown in Fig. \ref{ZA_graphs} b), the luminance of the cloudy night is higher than that of the clear night for almost all zenith angles. The clear sky has near-natural luminance values near zenith and the luminance rises towards the horizon to values of 2.5 mcd/m$^2$ at 86$^{\circ}$, before it drops sharply. The luminance of the cloudy sky is also close to 0.2 mcd/m$^2$ at zenith, remaining below 0.9 mcd/m$^2$ up to angles of 67$^{\circ}$. Above that angle, it rapidly rises before it peaks at 9.1 mcd/m$^2$ at about 83$^{\circ}$.

\subsubsection{Cloud attenuation factors}
We calculated the factor for brightening (amplification) or darkening (attenuation) due to the presence of clouds for the two investigated sky segments from the angular luminance distribution shown in Fig. \ref{ZA_graphs} a) and b). The result, obtained by dividing the clear sky data by the cloudy sky data, is plotted in Fig. \ref{ZA_graphs} c). Grey circles show the northeastern segment and black squares represent the sky segment facing the southwest. The attenuation by clouds observed in the southwestern segment (grey circles) is near unity at zenith rising slowly to values of about 3 at 65$^{\circ}$, with a local maximum of 4.8 at 37$^{\circ}$ due to the Milky Way. Above 65$^{\circ}$ it rises to a maximum of 11.3 at 86$^{\circ}$. In the other direction (north-east, black squares) the attenuation is also near unity at zenith and then drops slowly to values of about 0.5 at 70$^{\circ}$. Above 70$^{\circ}$ it drops to a minimum of 0.15 at 83$^{\circ}$, representing a brightening by a factor of about 6.7.

\subsubsection{Illuminance}
As described above, horizontal and scalar illuminance is calculated by the SQC software and it is further possible to estimate (a hypothetical) illuminance assuming a certain angular luminance distribution. In table \ref{table_ill_ZA}, the horizontal and scalar illuminance from the full image is provided and compared to the inferred illuminance values from the two selected sky segments, and for zenith luminance measured an SQM handheld photometer (see methods).
\begin{table}[h]
  \centering
  \caption{Illuminance values calculated from the angular luminance distribution for a clear and overcast night near Bonnievale, South Africa. (*) hypothetical value assuming an equal angular luminance distribution for all azimuth angles (**) inferred value assuming a constant luminance for all azimuth and zenith angles. Zenith luminance measured with DSLR and SQM are given as reference.}
  \label{table_ill_ZA}
  \begin{tabular}{ccc}
	\hline
&clear & clouds\\

horizontal Illuminance [mlx] \\
    \hline
average (full data) & 0.96 $\pm$ 0.10 & 0.76 $\pm$ 0.08\\
NE segment (*)  & 1.07 $\pm$ 0.11 & 2.55 $\pm$ 0.26\\
SW segment (*)  & 0.93 $\pm$ 0.09 & 0.34 $\pm$ 0.03\\
SQM (**) & 0.91 $\pm$ 0.10 & - \\
\hline
\\
scalar Illuminance [mlx] \\
    \hline
average (full data) & 1.98 $\pm$ 0.20 & 1.95 $\pm$ 0.20\\
NE segment (*)  & 3.20 $\pm$ 0.32 & 11.26 $\pm$ 1.13\\
SW segment (*)  & 1.77 $\pm$ 0.18 & 0.56 $\pm$ 0.06\\
SQM (**) & 1.82 $\pm$ 0.18 & - \\
\hline
\\
Zenith luminance [mcd/m$^2$]\\
\hline
SQM & 0.28 $\pm$ 0.03 & - \\
DSLR & 0.24 $\pm$ 0.03  & 0.19 $\pm$ 0.02 \\
  \end{tabular}
\end{table}

\section{Summary and Discussion}
\subsection*{Reference points for natural nocturnal light environments}
ALAN influences a large portion of the natural nocturnal landscapes worldwide \cite{falchi2016WA}, and its use is growing in area and intensity \cite{kyba2017VIIRS}. The NSB for clear skies is relatively well understood and a global model for NSB has been published recently \cite{falchi2016WA}. A natural clear starlit sky has a zenith luminance around 0.2-0.3 mcd/m$^2$ (depending on the position of the Milky Way and the amount of airglow), and that the clear sky horizontal illuminance is around 0.6-0.9 mlx \cite{haenel2018measuring}, setting lower limits for moonless clear nights.

While some studies regarding the brightening of the night sky by clouds in urban areas exist, very few data on the attenuation (darkening) by clouds has been published (but see \cite{Ribas2016clouds, Jechow2016, Posch2018} for data from single channel devices). A commonly cited value of 0.1 mlx for illuminance for cloudy nights \cite{biberman1971natural, seidelmann2005explanatory, book:rich_longcore} is based on extensive (daytime and nighttime) illuminance and luminance measurements performed by D.R.E. Brown at multiple locations several decades ago \cite{brown1952natural}. Brown's seminal report has produced excellent daylight and moonlight lookup charts that have been proven relatively robust. However, the accuracy of the Luckiesh-Taylor brightness meter \cite{luckiesh1937brightness} used for the nighttime measurements at very low light levels remains unclear. Furthermore, the attenuation by clouds at night were not measured but estimated. Knowledge of the lower limit of nocturnal darkness for cloudy nights, however, is an important reference point to estimate and model impacts of ALAN on species level or even for whole ecosystems. At this stage the bright values for cloudy nights in urban areas are compared to the reference values of a typical clear night (see e.g. \cite{Kyba:2015_isqm} for a brightening as large as a factor of 2200). However, cloudy night values obtained with ALAN should be better compared with the cloudy night values without ALAN as the amplification is much higher. The reference point idea can be put forward to other weather phenomena such as fog and extended to nights with moonlight, as darkening and brightening of moonlit skies will be much more complex and will depend on cloud height, type and optical thickness.

Reference points for illuminance under natural conditions are also useful for planning the experimental design of laboratory experiments, to mimic natural (daytime and nocturnal) conditions of specific habitats indoor. Right now, many laboratory experiments use simply ''darkness'' as a dark control level \cite{durrant2018artificial, bruening2018influence}  (a value of ''0 lx'' is often reported, although we note that this presumably indicates the illuminance was simply below sensor threshold, which in the case of inexpensive lightmeters could mean as much as 0.5 lx). Sometimes experiments use (often constant) moonlight levels on the order of 0.1-1 lx \cite{newman2015using} which does not represent a natural nocturnal light environment.

\subsection*{Cloud attenuation}
At the rural site at LakeLab in Germany, we observed an attenuation by clouds at zenith by a factor of 1.5 and a reduction of horizontal illuminance by a factor of 1.6, when taking into account the full data-set. Off-zenith cloud attenuation factors of more than 5 were observed for the sky segment least affected by ALAN and cloud amplification factors of more than 4 were obtained for the sky segment most affected by ALAN. The minimum off-zenith sky luminance was determined to be 0.11 mcd/m$^2$. From the sky segment least affected by skyglow, a hypothetical horizontal illuminance of 1.04 mlx for clear sky and 0.51 mlx for overcast sky was inferred from the measured angular luminance distribution. 

At the South African site, we observed an attenuation by clouds at zenith by a factor of 1.25, and a reduction of horizontal illuminance by a factor of 1.25. Off-zenith attenuation factors of more than 11 were obtained for the sky segment least affected by ALAN, while a brightening by a factor of more than 6 was observed in the opposite direction for the sky segment most affected by ALAN. The minimum off-zenith sky luminance was determined to be 0.05 mcd/m$^2$. By inspecting the sky segment least affected by skyglow, a hypothetical horizontal illuminance of 0.93 mlx for clear sky and 0.34 mlx for overcast sky was inferred. 

\subsection*{Color shift}
The multi-spectral information from the individual red, green and blue color channels of the camera indicates a spectral change with changing weather conditions. The CCT data shows an overall reduction (indicating a spectral shift towards red) in color temperature for overcast conditions at both sites, in agreement with previously reported results using SQMs and color filters in an urban context \cite{Kyba:2012_mssqm,aube2016spectral}. We are aware of the deficiencies of the CCT method to characterize artificial light sources that deviate from a perfect black body radiator \cite{valencia2013calibration, steigerwald2002illumination}. At this point, however, CCT is used widely in the lighting industry and the public awareness for CCT is growing. Furthermore, if the raw camera data is stored and archived properly, the three color channels can be investigated with a different method at a later stage. Please be aware that a standard DSLR is still limited for ecological studies as it lacks ultraviolet or near infrared sensitivity. A modification of digital cameras to extend the measurement range is not straightforward, but possible \cite{melis2011development}.

The ability to observe changes in the nighttime spectrum over time is important not only for astronomical observations, but also for the evaluation of possible ecological impacts of ALAN. Due to the transition towards solid-state lighting with LEDs, it is difficult to judge whether a site has become brighter or not \cite{kyba2018light} when devices are used that are not matching photometric visual band such as the SQM (see the comprehensive discussion on this topic by Sanchez de Miguel et al. \cite{sanchez2017sky} or recent work by Bouroussis \cite{bouroussis2018effect}) or even nighttime satellites, as the DMSP and VIIRS sensors lack sensitivity at short visible wavelengths \cite{Kyba:2015_viirs, kyba2017VIIRS}.

\subsection*{Distance shift}
The presence of clouds changes the ALAN sources that affect an individual site. While for clear skies distant ALAN sources can have a major impact on the NSB, these distant sources can be attenuated due to atmospheric conditions and local light sources can dominate with their contribution to the NSB. This effect will also depend on the cloud height \cite{Ribas2016clouds}. In our study, this effect is exemplified in the LakeLab data obtained near Berlin (best seen in the subtracted image Fig. \ref{LL_IMG} e). There, the long distance skyglow (e.g. from Berlin, Neustrelitz, Neubrandenburg) dominates the clear sky while local light sources (e.g. Rheinsberg and F{\"u}rstenberg) dominate the overcast sky. Most strikingly of the local light sources is the former nuclear power plant of Rheinsberg at ca. 2.5 km distance to the west. Due to the spatial information, we were able to identify the former power plant as the main source of (local) skyglow for the overcast conditions, while the distant skyglow from e.g. Berlin was attenuated by clouds. During the clear night, the light emission from the power plant and other local sites was not dominant in the NSB observations while the Berlin skyglow was prominent. However, the power plant light emission had only a small impact on the zenith luminance for both clear and cloudy night. This shift in source would not have been apparent via observations of zenith luminance with a single channel sensor devices such as an SQM or illuminance with a luxmeter. This means that a site that seems free from ALAN under conditions of clear skies (and therefore potentially fulfills requirements for a dark sky park \cite{kollath2017night}), is not necessarily free from ecological light pollution.

\subsection*{Advantages of DSLR photometry}
We have used a commercial DSLR camera with a fisheye lens for our study, as the easy-to-use method \cite{Kollath:2010,solano2016allsky,kollath2017night} has proven to be applicable for field work \cite{jechow2017measuring,jechow2017balaguer,jechow2018differential}. The data was processed with a commercial software, SkyQualityCamera, SQC (see methods) that has an intuitive  graphical user interface and a comprehensive set of features. SQC allows the analysis of such data without any programming skills or knowledge of photometry. An alternative to this software that requires some basic programming knowledge is the free software DiCaLum \cite{dicalum}, which we used in an earlier publication \cite{jechow2017balaguer}.

The all-sky multi-spectral imaging data is superior to single channel devices as the luminance maps not only provide a zenith NSB value, but also allow a comprehensive investigation of the angular luminance distribution and the color temperature. The spatially resolved data enables us to identify the (nearly) unpolluted parts of the sky and extract cloud attenuation factors for these parts, as well as to perform a direct comparison with polluted skies, using simultaneously acquired data. Although it is feasible to attach single channel devices to translation stages to scan the different segments of the sky \cite{zamorano2012nixnox} this is time consuming and requires very stable atmospheric conditions, which is problematic for cloud investigations.

Overall, the method applied here has potential to be widely used by amateurs as well as professionals from astronomy, life sciences or interdisciplinary fields to complement or replace SQM measurements.

\subsection*{Current obstacles and future directions}
At each site, two consecutive nights, one clear and one with clouds were investigated to ensure that artificial illumination and natural sky background are comparable. A comparison between different months or even seasons is not advisable, as lighting practice and natural background can change considerably. We note that it is rare to have the opportunity of being in an area that is almost unpolluted by ALAN, and then have the weather conditions change from clear to overcast from one night to the next, with both occurring during the monthly window of moonless nights. Despite this, we were able to acquire two data sets at two different rural locations. The amount of data available for further analysis could be increased either by setting up permanently measuring systems, or increasing the number of people acquiring data.

We therefore urge amateur and professional astronomers or interested citizen scientists as well as ALAN researchers from other fields to obtain such cloudy night data, and send it to light pollution researchers, or to analyze and publish it by themselves. A first step could be to approach us or the authors of the DiCaLum code \cite{dicalum}. Perhaps astrophotographers who travel long distances to very remote sites tend not to take images if the sky is overcast. However, such photographers could help building a data set that improves our understanding of light pollution. Furthermore, more of such studies on a systematic basis are required to cover different ecosystems (e.g. marine systems or polar regions), and different climate zones or other weather conditions like fog, snow and ice in the atmosphere.

\section{Conclusions}
Understanding of the Earth's illuminance on cloudy nights under natural conditions is a major knowledge gap. This reference point is essential for developing new experiments and performing ecological modeling in ALAN research. We have investigated the spectral color shift and the attenuation (or amplification) of the sky brightness by nocturnal clouds. Our main results can be listed as follows:

\begin{itemize}
\item Luminance levels as low as 0.05 mcd/m$^2$ (off-zenith) were observed for cloudy nights. Our results suggest an upper limit on natural illuminance under overcast conditions of 0.34 mlx. We expect more remote areas to have even lower illuminance.
\item Clouds strongly shift the distance scale of skyglow; sites that appear to be nearly unpolluted on clear nights might be quite strongly affected by (local) ALAN on cloudy nights.
\item A color shift to the red spectrum was observed for cloudy nights, which matches earlier observations in urban contexts \cite{Kyba:2012_mssqm,aube2016spectral}.
\end{itemize}

These results were obtained using a simple and robust method: differential all-sky photometry with a commercial digital single lens reflex camera and a fisheye lens. The results represent a first step towards discovering the missing reference points for natural nocturnal darkness in the context of ecological light pollution. Because both our sites were affected to some extent by artificial light, further investigations of the impact of clouds are necessary to fully understand the potential consequences of ecological light pollution on the environment. We urge that researchers and interested photographers not only measure zenith NSB on clear nights, but also use imaging sensors under different atmospheric conditions to quantify the nocturnal light environment.

\section{Methods}
\subsection{Study sites}
\subsubsection{LakeLab, Lake Stechlin, Germany}
The first set of observations were made at the LakeLab (53$^{\circ}$8$^{\prime}$36.5$^{\prime\prime}$N, 13$^{\circ}$1$^{\prime}$42.7$^{\prime\prime}$E), a floating ecological research platform installed in the south-western bay of Lake Stechlin, a clearwater lake with a surface area of 4.25 km$^2$ and a maximum depth of 69.5 m. The lake is situated about 75 km north of the city of Berlin in a forested area. The floating construction consists of one large enclosure with 30 m diameter and 24 smaller enclosures that are 9 m in diameter. The entire structure is anchored to the lake bottom to prevent drift. For more details about the LakeLab, see our previous report of the zenith brightness over several months \cite{Jechow2016}. The new World Atlas of artificial NSB \cite{falchi2016WA} predicts a clear sky luminance at zenith of 0.22 mcd/m$^2$ at this site \cite{falchi2016supplement}.
	\begin{figure}[h]
		\centering
\includegraphics[width=0.8\columnwidth]{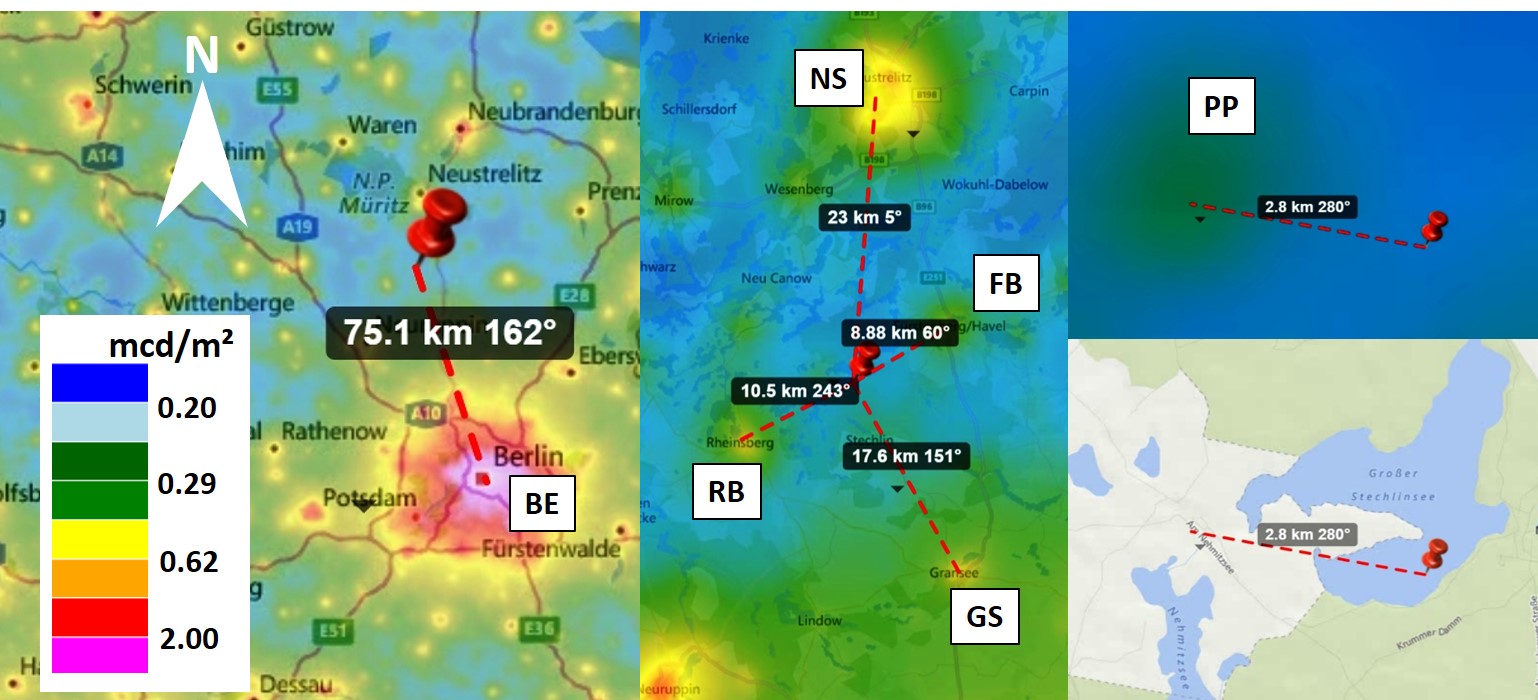}
\caption{Map of the Lake Stechlin and Berlin area indicating the location of the LakeLab and nearby settlements. BE-Berlin, NS-Neustrelitz, FB-F{\"u}rstenberg, RB-Rheinsberg, GS-Gransee, PP-power plant.  The false color overlay shows calculated NSB on clear nights from \cite{falchi2016supplement}. Map data from www.lightpollutionmap.info by Jurij Stare.}
\label{WA_LL}
\end{figure}

Lake Stechlin is part of a nature reserve belonging to one of the areas of Germany least affected by skyglow (according to Falchi et al. \cite{falchi2016WA}). The nearest village is Neuglobsow (53$^{\circ}$8$^{\prime}$48$^{\prime\prime}$N, 13$^{\circ}$2$^{\prime}$49$^{\prime\prime}$E) with less than 300 inhabitants and the largest nearby major settlement is the city of Berlin (distance ca. 75 km, azimuth ca. 162$^{\circ}$, about 3.5 Mio. inhabitants) in the south-south-east. Other settlements contributing to skyglow at LakeLab are the towns of Neustrelitz (ca. 25 km distance, azimuth ca. 5$^{\circ}$, ca. 20,000 inhabitants) and Neubrandenburg (ca. 50 km distance, azimuth ca. 10$^{\circ}$, ca. 65,000 inhabitants), both to the north, F{\"u}rstenberg (ca. 9km distance, azimuth ca. 60$^{\circ}$, ca. 6000 inhabitants) in the northeast and Rheinsberg (ca. 10km distance, azimuth ca. 243$^{\circ}$, ca. 8000 inhabitants) in the southwest. The former nuclear power plant of Rheinsberg is located at ca. 2.8 km distance to the west (azimuth ca. 280$^{\circ}$). For locations and distances see maps in Fig. \ref{WA_LL}.

\begin{figure}[tp]
\centering
\includegraphics[width=0.6\columnwidth]{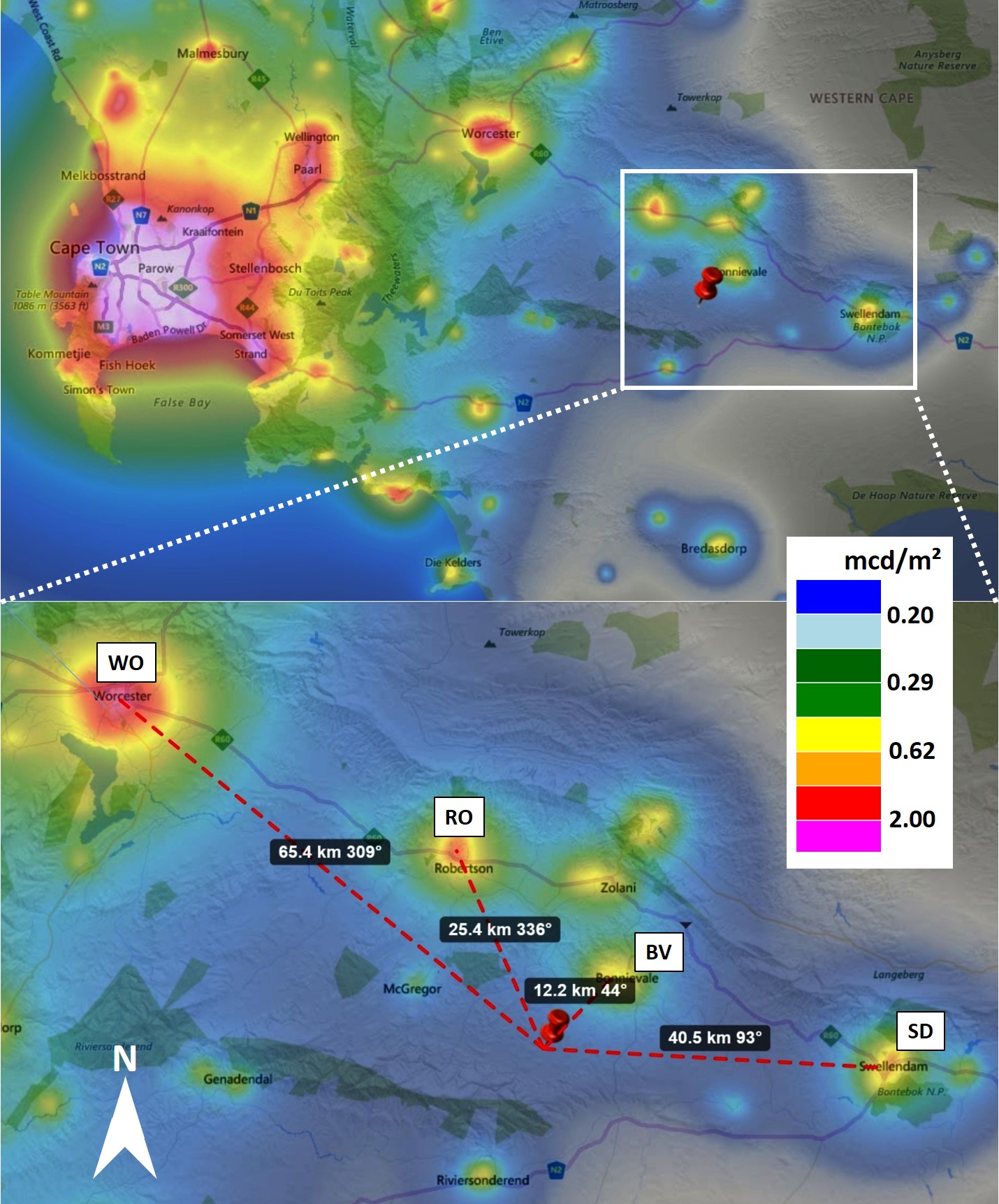}
\caption{Map of the Cape Town, Bonnievale area in South Africa. BV-Bonnievale, RO-Robertson, WO-Worcester, SD-Swellendam. The false color overlay shows calculated NSB from \cite{falchi2016supplement}. Map data from www.lightpollutionmap.info by Jurij Stare.}
\label{WA_ZA}
\end{figure}

\subsubsection{Bonnievale, South Africa}
The second set of observations were made at the Night Sky Caravan Park (34$^{\circ}$0$^{\prime}$44.8$^{\prime\prime}$S, 19$^{\circ}$59$^{\prime}$41.4$^{\prime\prime}$E), an astro-tourism site near Bonnievale, South Africa. The site lies in the Cape Winelands district, part of the western Cape province about 180 km east of Cape Town, see Fig. \ref{WA_ZA} for a map of the region. Nearby settlements are the towns of Bonnievale (12 km distance, azimuth ca. 12$^{\circ}$, ca. 9000 inhabitants) to the northeast, Robertson (25 km, azimuth ca. 335$^{\circ}$, ca. 28.000 inhabitants) to the north, Swellendam (40 km, azimuth ca. 90$^{\circ}$, 17.000 inhabitants) to the east and Worcester (65 km, azimuth ca. 310$^{\circ}$, ca. 97.000 inhabitants) to the northwest.

\subsection{DSLR camera and image processing software}
Images were obtained with a commercial DSLR camera, a Canon EOS 6D, which has a full-frame 20.2 Megapixel (5496$\times$3670) CMOS sensor. The camera was mounted on a tripod and operated with a circular fisheye lens (Sigma EX DG with 8 mm focal length) always at full aperture of 3.5. To acquire all-sky images, the camera was aligned with the center of the lens oriented towards the zenith.  At Lake Stechlin, Germany, ISO 3200 and shutter speed of 30 s was used. At Bonnievale in South Africa, the camera was set to ISO 1600 and a shutter speed of 30 s.

For image processing we used the commercial ''Sky Quality Camera'' software (Version 1.8, Euromix, Ljubljana, Slovenia). The camera is calibrated by the software manufacturer. The manufacturer's calibration includes photometric calibration using the green channel of the camera, as well as correction of optical aberrations like vignetting. The software processes the luminance $L_{v,sky}$ of the sky for each pixel of the camera (luminance is commonly referred to as ''brightness'' referenced to human vision). The software calculates the cosine corrected illuminance $E_{v,cos}$ in the imaging plane:
\begin{equation}
E_{v,cos}=\int_{0}^{\pi \over 2}\int_{0}^{2\pi} L_{v,sky}(\theta, \phi) sin\theta cos\theta d\phi d\theta,
\end{equation}
and the scalar illuminance for the imaging hemisphere $E_{v,scal,hem}$ without cosine correction \cite{duriscoe2016photometric}:
\begin{equation}
E_{v,scal,hem}=\int_{0}^{\pi \over 2}\int_{0}^{2\pi} L_{v,sky}(\theta, \phi) sin\theta d\phi d\theta.
\end{equation}
In the equations, $L_{v,sky}$ is the sky luminance, $\theta$ is the zenith angle and $\phi$ is the azimuth angle. For all-sky images, i.e. when imaging in the horizontal plane, $E_{v, cos}$ is usually termed horizontal illuminance.

From the three spectral channels, color correlated temperature (CCT) is calculated by the software. CCT describes the temperature of a thermal source that most closely matches the color spectrum of the observed light. Candle light has a low CCT (around 2000 K), the sun has a CCT of about 5800 K, and even higher temperatures appear bluish-white. In urban areas, a spectral shift from blue to red (i.e. reduction in color temperature) has been reported for overcast compared to clear skies \cite{Kyba:2012_mssqm,aube2016spectral}. The software provides luminance and CCT maps, allows the subtraction of one image from another, to analyze sectors of the image, and to do cross sections. 

\subsection{Sky quality meter (SQM)}
For comparison to the imaging results, we used a sky quality meter with a lens (SQM-LU). It measures the sky radiance $L_{\mathrm{SQM}}$ in units $mag_{\mathrm{SQM}}/arcsec^2$ in a cone with a full width half maximum of 20$^{\circ}$ in a single spectral channel only approximately resembling human photopic vision (see review comparing devices \cite{haenel2018measuring}, or \cite{sanchez2017sky} for a detailed discussion of the spectral sensitivity of the SQM). Due to the spectral mismatch, the sky radiance $L_{\mathrm{SQM}}$ measured with an SQM can only approximately be converted to a sky luminance at zenith $L_{v,sky,zen}$ using:

\begin{equation}
\label{eq:cdm2_V}
L_{v,sky,zen} \, [\mathrm{cd/m}^2] \, \approx \, 10.8 \times 10^4 \times 10^{-0.4*L_{\mathrm{SQM}}}.
\end{equation}

Please be aware that this commonly used equation overestimates the precision achievable with the SQM. The interested reader is referred to some recent work by Bar{\'a} on the conversion \cite{bara2017variations}. Using SQM data obtained at zenith, we inferred illuminance values assuming a homogeneous brightness distribution over the hemisphere according to $E_{v,hor} \approx \pi \cdot  L_{v,sky,zen}$ \cite{kocifaj2015zenith}, which is a rough estimation commonly done in the literature to link SQM measurements with illuminance measurements often done in field work. While the angular NSB distribution will vary from site to site and from season to season, this crude estimate provides a reasonable approximation. A discussion on the relation between zenith sky brightness and horizontal illuminance is outside the scope of this work, but can be found in \cite{kocifaj2015zenith}.

\subsection{Estimating cloud base height from the all-sky imaging data}
The information of cloud base height is important for studies of ALAN on overcast nights. Ribas et al. used laser ceilometer data in their study \cite{Ribas2016clouds}, but such data is not always easily accessible. With the all-sky imaging technique used here, the cloud base height information can be inferred from the imaging data itself. If the distance to a light source is known it is possible to estimate the cloud base height $h_{CB}$ from the projection of the light on the cloud base by triangulation using the distance $d$ and the elevation angle $\epsilon$, with $tan(\epsilon)=h_{CB}/d$. This method was used in the past to estimate cloud heights in optical ceilometers using a narrow beam of light \cite{ashford1947} and recently demonstrated with night-time imaging of light pollution \cite{Kollath2016ALAN}. For the German site at LakeLab, the distance to the power plant is ca. 2.8 km with $\epsilon \approx 17^{\circ} \pm 1^{\circ}$ resulting in a cloud base height of $h_{CB} \approx$ 850 m $\pm$ 100 m. This value can be confirmed by checking data for F{\"u}rstenberg. The distance here is ca. 8.9 km with $\epsilon \approx 6^{\circ} \pm 1^{\circ}$ resulting in a cloud base height of $h_{CB} \approx$ 950 m $\pm$ 150 m. For the South African site, the distance to the town of Bonnievale is ca. 12 km with $\epsilon \approx 7^{\circ} \pm 1^{\circ}$, which results in a cloud base height of $h_{CB} \approx$ 1.5 km $\pm 0.2$ km. Please be aware that this method does not work at large distances due to the curvature of the Earth and lens aberrations for small angles.

\section*{Author contribution}
AJ conceived the study with input from FH and CCMK. CCMK acquired the South Africa data. AJ acquired the Lake Stechlin data, analyzed all data and wrote the first draft of the manuscript. The final manuscript was written with contributions from all authors.

\section*{Competing interests}
The authors declare no competing interests.

\section*{Funding}
This work is funded by the Leibniz Association, Germany (SAW-2015-IGB-1) within the ILES (Illuminating Lake EcoSystems) project. We further acknowledge funding by the COST (European Cooperation in Science and Technology) Action ES1204 Loss of the Night Network (LoNNe) for supporting a European network of light pollution researchers that inspired this work. CCMK acknowledges funding from the European Union's Horizon 2020 research and innovation programme under grant agreement no. 689443 via project GEOEssential.

\section*{Acknowledgements}
We thank the ILES team for moral support, Zolt{\'a}n Koll{\'a}th for essential discussions on the topic of ALAN and photometry with DSLR cameras. CCMK deeply thanks Auke Slotegraaf, Case Rijsdijk, Ed and Lynnette Foster, and the members of the Astronomical Society of southern Africa for their invitation to visit South Africa and their warm hospitality during his stay. AJ thanks Manuel Spitschan and Travis Longcore for help in literature research.

\section*{Data accessibility}
The datasets generated during and/or analysed during the current study are available from the corresponding author on request.

\end{document}